\documentclass[11pt]{article}

\def\references#1{\vspace*{5mm}\noindent{References:}\list
{[\arabic{enumi}]}{\settowidth\labelwidth{[#1]}\leftmargin\labelwidth
\advance\leftmargin\labelsep
\usecounter{enumi}}
\def\newblock{\hskip .11em plus .33em minus .07em}
\sloppy\clubpenalty4000\widowpenalty4000
\sfcode`\.=1000\relax}

\begin{document}

\setcounter{page}{1}

\begin{center}
{\large \sf
	
	A Farewell to Professor RNDr. V\v{e}ra Trnkov\'a}, DrSc.
 
% \setcounter{footnote}{0}\footnote{A joint work with J. Rosick\'{y}}

%Technical University of Braunschweig

\vspace*{5mm}
{\large \em  J. Ad\'{a}mek}
\end{center}

%%%%%%%%%
When in May 2018 V\v{e}ra Trnkov\'a passed away, the mathematical community lost a
marvellous personality who had substantially influenced the generation in which she 
was active as a scientist and teacher. Her lectures were full of her enthusiasm and 
love for math. And she led her seminar in a wonderfully inspiring manner for almost
half a century -- when her disease started to prevent her going to the university, 
it was held in her flat. Her radiant personality and her great enthusiasm made
it a pleasure to collaborate with her.

V\v{e}ra Trnkov\'a was born in 1934, she studied mathematics at the Charles University in Prague
in 1952-1957 and continued there,  at the Faculty of Mathematics an Physics, with her doctoral studies supervised by Professor Eduard \v{C}ech
in 1957-1960. From 1960 until her emeritation she was employed at that faculty. In 1961 she obtained title CSc (corresponding to PhD). For political reasons she had to wait for reaching the higher scientific degree
DrSc until 1989 and the Full Professor position until 1991.

Besides research, she had a  great love for the forest. Her father was a forester, and she
passionately collected mushrooms and antlers (of which the walls of her study were full).
She dreamed of hunting, but never made any practical steps
in that direction. In her last couple of years she gave up research and wrote short stories 
and two volumes of personal memories. V\v era could also make nice drawings, e.g. the 
above cat symbol for the conference Categorical Topology held in 1988 in Prague. 

 V\v{e}ra had a very wide scope of interests in research, from General Topology
through Category Theory to Theoretical Computer Science. I will try to indicate some of her best
results. But I admit that my choice will be a substantial reduction of all that V\v{e}ra
achieved in her more than 160 reseach articles and the two monographs she co-authored.

The first papers of V\v{e}ra's were devoted to General Topology
and were much influenced by Profesor Miroslav Kat\v{e}tov. Surprisingly,
the most widely cited topological article of V\v{e}ra's is her very first publication (in Russian and 
published under her maiden name \v{S}ediv\' a), based on her master thesis, see
[11]. General Topology remained her great love all through her scientific career.
For example, a series of her publications applied a very special continuum (a connected, compact Hausdorff
space) constructed by H. Cook in 1967 and sketched on four pages of his article [3]. When V\v{e}ra decided 
ten years later to unravel all the details of Cook's construction, she virtually spent several months doing nothing else. 
As a result, she presented a detailed proof of fifty(!) tightly written pages in her joint monograph [7].
Using Cook's continuum, V\v{e}ra proved a number of really astonishing results. For example:
every monoid is isomorphic to the monoid of all non-constant continuous  self-maps of a 
regular space on which all continuous real functions are constant [16].

V\v{e}ra was in the late 1950's one of the first mathematicians in Czechoslovakia to recognize the importance
of category theory. Her first papers in that realm were devoted to formal completions of categories. But soon her
interest shifted to embeddings of concrete catgeories (i.e., those 
with a faithful functor to $\emph Set$). This was influenced by the research reported by Zden\v{e}k Hedrl\'{i}n and Ale\v{s} Pultr
at the Topology Seminar (led by Miroslav Kat\v{e}tov), see [4]. The question was which concrete category is 'universal'
in the sense that every concrete category can be fully embedded into it? 
 For example, the category of topological spaces and local homeomorphisms and that of unary algebras on two operaions 
 were proved to be universal, assuming certain set-theoretical assumptions. Later
 V\v era concentrated on embeddings into the category of topological spaces and continuous 
 maps.  Here the fact that constant maps are always
 morphisms led her to introducing the \emph {almost universal}
 categories $\mathcal K$: this means that every concrete category has an almost full embedding
 into $\mathcal K$. That is, an embedding $E$ such that $Ef$ is
 non-constant for every morphism $f$ and, conversely, every non-constant
 morphism between objects of  $E[\mathcal K]$ has the form $Ef$ for a
 unique $f$. V\v{e}ra proved in [14] that the category of compact Hausdorff spaces
 is, under some set-theoretical assumptions, almost universal.
 (Without any set-theoretical assumptions all algebraic categories have an almost full embedding into the category
 of paracompact Hausdorff spaces, as one of the most brilliant doctoral students of V\v{e}ra's, V\'aclav Koubek,
 proved [5].) In V\v{e}ra's joint monograph with Ale\v{s} Pultr [7] an impressive
 theory of embeddings of concrete categories is presented.
 
 Another passion of V\v{e}ra's were set functors. A beautiful theory of their
 properties was presented by her in the early 1970's, see [12, 13], and she
 continued studying and applying them throughout the following decades. One
 of her results that has been particularly often used recently is
 that for every set functor $F$ there exists a
 functor $G$ preserving finite intersections that agrees
 with $F$ on all non-empty sets. Nowadays
 $G$ is called the \emph{Trnkov\'a hull} of $F$.  V\v{e}ra and members of her
 seminar applied her insight to the theory of algebras and automata for
 a given set functor (or, more generally, an endofunctor of a category where the
 underlying objects 'live'). V\v{e}ra's beautiful proof that a set functor $F$ generates a free monad 
 iff it has arbitrarily large prefixed points (i.e.~sets $X$ of cardinality
larger or equal to that of $FX$) was generalized to arbitrary endofunctors in [20].
An important, often cited paper [18] of V\v{e}ra's concerns the behaviour
of nondeterministic automata. Results obtained in V\v{e}ra's seminar on
automata and algebras for endofunctors are collected in our joint monograph [1].
I remember with pleasure the time we worked on this book; we often had fierce
discussions, but invariably we would find by the next day a solution that we both found good.
Jan Reiterman, another of the brilliant doctoral students of V\v{e}ra, devoted his
dissertation to iterative constructions of algebras. One of his results, known these days as the Reiterman Theorem,
is widely used in the Theoretical Computer Science: it characterizes classes
of finite algebras that are pseudovarieties, i.e.~closed under subalgebras, 
regular quotients and finite products, see [8]. I remember how happy 
V\v{e}ra was when Jan presented this result in her seminar.

 A realm in which V\v{e}ra also achieved astounding results is that of isomorphisms
 of products of algebras and spaces. For example in [19] she constructed a metric space $X$
 homeomorphic to its power $X^2$ and uniformly homeomorphic  to $X^4$
 but not to $X^2$ and at the same time isometric to $X^8$ but not to $X^4$.
 A famous result of hers is that whenever a countable Boolean algebra $B$ is isomoprhic to
 the coproduct $B+B+B$, it is also isomorphic to $B+B$. Her paper [17] was reprinted in the Handbook of
 Boolean Algebras [6] where R.S. Pierce comments: 'It was a major surprise when Trnkov\'a showed
 that the multiplicative analog of the cube problem has a positive solution in BA'.
She proved with coauthors that every finite abelian group has a representation 
by both products and coproducts of boolean algebras. That is,
a collection $B_x$ of pairwise non-isomorphic boolean algebras is given, indexed by
elements $x$ of the group, with $B_{x+y}$ always isomorphic  to
$B_x \times B_y$ (see [2]) or to $B_x + B_y$ (see [21]), respectively.
For \emph{all} commutative semigroups  V\v{e}ra proved such a representation by products
of metric spaces in [19]. There was in fact
much much more V\v{e}ra proved about such representations, including a series of
articles on clones represented by products in various categories. A number of these results
were obtained in collaboration with Ji\v{r}\'{i} Sichler, see e.g.~[9,10].

V\v{e}ra had at least eight 'doctoral children' (I do not use the expression PhD students 
since the title was called CSc, a candidate of sciences, until the 1990s).  In the current issue
they are represented by Libor Barto and myself. Two names that are sorely missing are those
of Koubek and Reiterman: they passed away before V\v{e}ra. It seems impossible to find
out even a good approximation of the number of her doctoral descendants; her 'doctoral grandchildern'
are represented in the present issue by my PhD students Lurdes Sousa and Stefan Milius, and her 'great-grandchildern'
by Thorsten Wi\ss mann, a PhD student of Milius. V\v{e}ra had close contacts with the group
of General Topology in Moscow, represented here by Alexander Vladimirovich Archangel'skii, and with a number of 
groups all over the world. I am happy  about the number of colleagues and friends who
have contributed to this volume.

I was entrusted to organize the refereeing process for the articles in the present volume. The article I co-author followed the standard submission procedure of the journal. I am grateful to the Executive Editor Zdenka Crkalov\' a for her very pleasant cooperation.

\vspace {5mm} 
%%%%%%%%%
REFERENCES

\medskip

[1] J. Ad\'amek and V. Trnkov\'a, Automata and algebras in a category, Kluwer
Academic Publ., Dordrecht 1989

[2] J. Ad\'amek, V. Koubek and V. Trnkov\'a, Sums of boolean spaces represent
every group, Pacific J. Mathem. 61 (1975), 1-6

[3] H. Cook, Continua which admit only the identity mapping onto non-degenerate
subcontinua, Fund. Math. 69 (1967), 241-249

[4] Z. Hedrl\'in, A. Pultr and V. Trnkov\'a, Concerning a categorical approach to topological and algebraic
theories, Proceedings of the Second Prague Topological Symposium, Prague 1966, 176-181

[5] V. Koubek, Each concrete category has a representation by $T_2$-paracomact spaces,
Comment. Math. Univ. Carolinae 15 (1974), 655-663

 [6] J. D. Monk and R. Bonnet (ed.), Handbook of Boolean algebras, North Holland 1988
 
 [7] A. Pultr and V. Trnkov\'a, Cominatorial, algebraic and topological representations of
 groups, semigroups and categories, Academia Praha 1980
 
 [8] J. Reiterman, The Birkhoff theorem for finite algebras, Algebra Univeralis 14 (1982),
 1-10
 
 [9] J. Sichler and V. Trnkov\'{a}, Disciplined spaces and centralizer clone segments, 
 Canad. J. Math 48 (1996), 1296-1323

 [10] J. Sichler and V. Trnkov\'{a}, Continuous maps of products of metrizable spaces,
 Houston J. Math. 26 (2000), 417-450
 
 [11] V. \v{S}ediv\'a, On collectionwise normal and hypercompact spaces (in Russian), 
 Czechoslovak Math. J. 9 (1959), 50-62
 
 [12] V. Tnkov\'a, Some properties of set functors,  Comment. Math. Univ. Carolinae
 10 (1969), 323-352
 
 [13]  V. Tnkov\'a, Descriptive classification of set functors I, II,
 Comment. Math. Univ. Carolinae 12 (1971), 143-175 and 345-357
 
[14] V. Trnkov\'a, Non-constant continuous mappings of metric or compact
Hausdorff spaces, Comment. Math. Univ. Carolinae 13 (1972), 283-295

[15] V. Trnkov\'a, Representation of semigroups by products in a category, J. of Algebra 34 (1975), 191-204

[16] V. Trnkov\'a,  Categorical aspects are useful in topology, Lecture Notes Mathem. 609, Springer Verlag
1977, 211-225

[17] V. Trnkov\'a, Isomorphisms of sums of countable Boolean algebras, Proc. Amer. Math. Soc. 80 (1980)
389-392

[18] V. Trnkov\'a, General theory of relational automata, Fund. Informaticae 3 (1980)
189-233

[19] V. Trnkov\'{a}, Products of metric, uniform and topological spaces,
Comment. Math. Univ. Carolinae 31 (1990), 167-180 

[20] V. Trnkov\'a, J. Ad\'amek, V. Koubek and J. Reiterman, Free algebras, input
processes and free monads, Comment. Math. Univ. Carolinae 16 (1975), 339-351
 
[21] V. Trnkov\'a and V. Koubek, Isomorphisms of sums of Boolean algebras, 
Proc. Amer. Math. Soc. 66 (1977), 231-236

\end{document}